\documentclass[5p,twocolumn,times,number]{elsarticle}

\usepackage{graphicx}
\usepackage{amsmath}   
\usepackage{microtype}   

\begin{document}

\begin{frontmatter}

\title{The Tunka Radio Extension (Tunka-Rex): Radio Measurements of Cosmic Rays in Siberia}

\author[add2]{F.G.~Schr\"oder}\corref{cor}
\ead{frank.schroeder@kit.edu}
\author[add1]{P.A.~Bezyazeekov}
\author[add1]{N.M.~Budnev}
\author[add1]{O.A.~Gress}
\author[add2]{A.~Haungs}
\author[add2]{R.~Hiller}
\author[add2]{T.~Huege}
\author[add1]{Y.~Kazarina}
\author[add3]{M.~Kleifges}
\author[add1]{E.N.~Konstantinov}
\author[add4]{E.E.~Korosteleva}
\author[add2]{D.~Kostunin}
\author[add3]{O.~Kr\"omer}
\author[add4]{L.A.~Kuzmichev}
\author[add4]{N.~Lubsandorzhiev}
\author[add1]{R.R.~Mirgazov}
\author[add1]{R.~Monkhoev}
\author[add1]{A.~Pakhorukov}
\author[add1]{L.~Pankov}
\author[add4]{V.V.~Prosin}
\author[add5]{G.I.~Rubtsov}
\author[add6]{R.~Wischnewski}
\author[add1]{A.~Zagorodnikov} 

\address[add2]{Institut f\"ur Kernphysik, Karlsruhe Institute of Technology (KIT), Germany}
\address[add1]{Institute of Applied Physics ISU, Irkutsk, Russia}
\address[add3]{Institut f\"ur Prozessdatenverarbeitung und Elektronik, Karlsruhe Institute of Technology (KIT), Germany}
\address[add4]{Skobeltsyn Institute of Nuclear Physics MSU, Moscow, Russia}
\address[add5]{Institute for Nuclear Research of the Russian Academy of Sciences, Moscow, Russia}
\address[add6]{DESY, Zeuthen, Germany}

\cortext[cor]{Corresponding author}

\begin{abstract}
The Tunka observatory is located close to Lake Baikal in Siberia, Russia. Its main detector, Tunka-133, is an array of photomultipliers measuring Cherenkov light of air showers initiated by cosmic rays in the energy range of approximately $10^{16}-10^{18}\,$eV. In the last years, several extensions have been built at the Tunka site, e.g., a scintillator array named Tunka-Grande, a sophisticated air-Cherenkov-detector prototype named HiSCORE, and the radio extension Tunka-Rex. Tunka-Rex started operation in October 2012 and currently features 44 antennas distributed over an area of about $3\,$km\textsuperscript{2}, which measure the radio emission of the same air showers detected by Tunka-133 and Tunka-Grande. Tunka-Rex is a technological demonstrator that the radio technique can provide an economic extension of existing air-shower arrays. The main scientific goal is the cross-calibration with the air-Cherenkov measurements. By this cross-calibration, the precision for the reconstruction of the energy and 
mass of the primary 
cosmic-ray particles can be determined. Finally, Tunka-Rex can be used for cosmic-ray physics at energies close to $1\,$EeV, where the standard Tunka-133 analysis is limited by statistics. In contrast to the air-Cherenkov measurements, radio measurements are not limited to dark, clear nights and can provide an order of magnitude larger exposure.
\end{abstract}


\end{frontmatter}

\section{Introduction}
The amplitude of the radio emission by cosmic-ray air-showers is roughly proportional to the energy of the primary particle \cite{FalckeNature2005}. This calorimetric sensitivity yields complementary information compared to the detection of secondary air-shower particles on ground. During dark nights with clear sky similar complementary information is available by measurements of air-Cherenkov and air-fluorescence light. Despite the higher energy threshold of about $10^{17}\,$eV, the radio technique would be and interesting alternative due to its full-time availability, provided that the cost and precision is at least comparable.

\begin{figure}[t]
  \centering
  \includegraphics[width=0.91\linewidth]{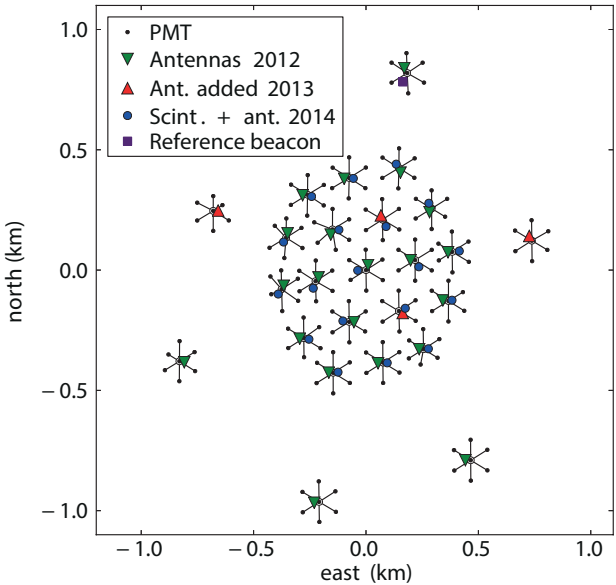}
  \caption{Map of the detector arrays for cosmic-ray air showers at the Tunka site: the antenna array Tunka-Rex at different stages, the air-Cherenkov array Tunka-133, and the scintillator array Tunka-Grande.}
  \label{fig_map}
\end{figure}

The main scientific goal of Tunka-Rex is the cross-calibration of radio and air-Cherenkov measurements of the same air showers, to test the achievable precision and real potential of the radio technique experimentally. For this purpose, Tunka-Rex is built as extension of the Tunka-133 photomultiplier array in Siberia \cite{Tunka133_NIM2014} (Fig.~\ref{fig_map}), which measures the air-Cherenkov light of showers in the energy range of approximately $10^{16}-10^{18}\,$eV. The antennas are read out in parallel with the air-Cherenkov detector which provides the necessary hybrid measurements for cross-calibration.

\section{Detector description}
Since the radio emission of air showers is strongest at wavelengths of a few meters, Tunka-Rex measures in the effective band of $35-76\,$MHz, similar to other antenna arrays, like LOPES \cite{FalckeNature2005}, CODALEMA \cite{ArdouinBelletoileCharrier2005}, LOFAR \cite{SchellartLOFAR2013}, or AERA \cite{AERA_PISA2015}. In fact, large parts of the analog electronics and even the Offline analysis software \cite{RadioOffline2011} have originally been developed for AERA, and then optimized for the conditions of Tunka-Rex.

In contrast to other experiments Tunka-Rex uses SALLA antennas \cite{AERA_AntennaPaper2012}. On the one hand, they feature a relatively low gain, causing a slightly higher detection threshold. On the other hand, the gain is almost independent of changing ground conditions which guarantees low systematic uncertainties. The antennas, including the whole signal chain, have been calibrated 
with a reference source used already by LOPES \cite{NehlsCalibration}, as well as laboratory measurements of individual parts \cite{TunkaRex_NIM2015}. Thus, Tunka-Rex results can be compared on an absolute level to other experiments and to theoretical predictions.

Time calibration is important for several reasons, not only for the reconstruction of the shower direction. A relative timing accuracy of a few nanoseconds within each event is necessary to reconstruct the shape of the radio wavefront containing information on the position of the shower maximum \cite{LOPES_Wavefront2014}, which itself is a statistical estimator for the mass composition of the primary particles. If the time synchronization is stable to a level of $1\,$ns, interferometric methods could be used, lowering the detection threshold \cite{FalckeNature2005}. 

\begin{figure}[t]
  \centering
  \includegraphics[width=0.99\linewidth]{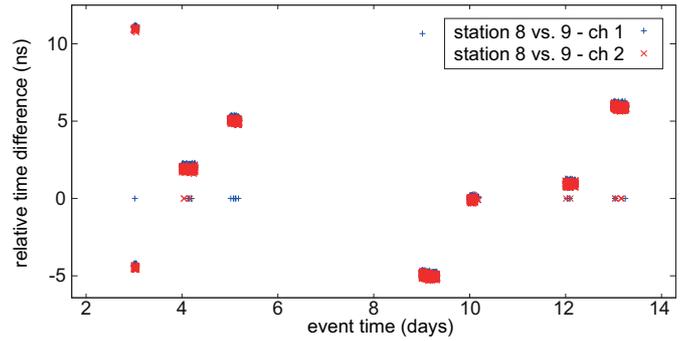}
  \caption{Relative timing between two Tunka-Rex antenna stations located close to the reference beacon over several nights of measurements. Each point denotes the time difference between both stations for one event. For a few events the beacon analysis failed (points exactly at $0\,$ns), and in the first night the beacon analysis did not converge to a stable solution (due to intrinsic phase ambiguities in the beacon analysis, i.e, the solutions correspond to $\pm 2\pi$). The evaluation of the beacon phases with both antenna channels yields consistent results. In a few nights measurements are missing due to bad weather.}
  \label{fig_timing}
\end{figure}

The time synchronization of Tunka-Rex is provided by the Tunka-133 host experiment and it has been checked with a reference beacon designed for the LOPES experiment \cite{SchroederTimeCalibration2010}. This beacon emits continuous sine waves whose phasing can be used to monitor possible time variations between the stations. The beacon has been used sporadically for a few nights (Fig.~\ref{fig_timing}). During each night the time synchronization is stable to sub-nanosecond level. However, in between nights, the detector is re-initialized and jumps of a few nanoseconds occur. We will investigate to which extent these jumps can be corrected in the later offline analyses. Such a correction could enable analyzing the wavefront shape and using digital interferometry.

\section{First results and outlook}
Already in the first season of operation (October 2012 - April 2013), Tunka-Rex successfully detected air-shower events. As expected, the measured amplitude depends on the shower energy, and on the angle between the geomagnetic field and the shower direction \cite{TunkaRex_RICAP2013}. Moreover, we have found indications that the shape of the radio lateral distribution is correlated with the atmospheric depth of the shower maximum, $X_\mathrm{max}$, reconstructed by the air-Cherenkov measurements. This sensitivity is expected theoretically and indicated by other experimental results \cite{LOPES_PRD2012, LOFAR_Xmax2014}, but has not yet been confirmed purely experimentally. The second season data (October 2013 - April 2014) will be used for a blind test of the result. Since the results of the air-Cherenkov measurements are blinded, a prediction of the energy and the shower maximum based on the radio measurements can be used as ultimate cross-check of the Tunka-Rex 
precision for these observables. By cross-calibration the absolute scale of Tunka-Rex can be linked to the presently more accurate scale of Tunka-133, such that the total expected accuracy of Tunka-Rex is equal to those of Tunka-133.

Provided that the unblinded data confirm that the Tunka-Rex precision is indeed comparable to the Tunka-133 precision, then the radio technique qualifies for its application to cosmic-ray science. For this purpose, Tunka-Rex has already been extended by deploying additional antennas, which will be triggered by the new scintillator array Tunka-Grande at the same site \cite{TAIGA_ECRS2014}. Using the scintillators as trigger during day-time and bad-weather periods, Tunka-Rex can significantly increase the statistics around $10^{18}\,$eV, which is exactly the energy range for which the present Tunka-133 analyses is limited by statistics.

\subsection*{Acknowledgement}
\footnotesize{Tunka-Rex has been funded by the German Helmholtz association and the Russian Foundation for Basic Research (grant HRJRG-303). Moreover, this work was supported by the Helmholtz Alliance for Astroparticle Physics (HAP), by the Russian Federation Ministry of Education and Science (agreement 14.B25.31.0010, zadanie 3.889.2014/K), the Russian Foundation for Basic Research (Grants 12-02-91323, 13-02-00214, 13-02-12095, 14002-10002) and the President of the Russian Federation (grant MK-1170.2013.2).}


\end{document}